\documentclass[aps,prb,twocolumn,superscriptaddress]{revtex4-1}
\usepackage{graphicx}
\begin{document}


\title{Femtosecond phase-resolved microscopy of plasmon dynamics in individual gold nanospheres}


\author{Francesco Masia}
\affiliation{Cardiff University School of Physics and Astronomy, The
Parade, Cardiff CF24 3AA, United Kingdom}\affiliation{Cardiff
University School of Biosciences, Museum Avenue, Cardiff CF10 3AX,
United Kingdom}
\author{Wolfgang Langbein}
\affiliation{Cardiff University School of Physics and Astronomy, The
Parade, Cardiff CF24 3AA, United Kingdom}
\author{Paola Borri}
\email{borrip@cardiff.ac.uk}\affiliation{Cardiff University School
of Biosciences, Museum Avenue, Cardiff CF10 3AX, United
Kingdom}\affiliation{Cardiff University School of Physics and
Astronomy, The Parade, Cardiff CF24 3AA, United Kingdom}



%
\begin{abstract}
The selective optical detection of individual metallic nanoparticles
(NPs) with high spatial and temporal resolution is a challenging
endeavour, yet is key to the understanding of their optical response
and their exploitation in applications from miniaturised
optoelectronics and sensors to medical diagnostics and therapeutics.
However, only few reports on ultrafast pump-probe spectroscopy on
single small metallic NPs are available to date. Here, we
demonstrate a novel phase-sensitive four-wave mixing (FWM)
microscopy in heterodyne detection to resolve for the first time the
ultrafast changes of real and imaginary part of the dielectric
function of single small ($<40$\,nm) spherical gold NPs. The results
are quantitatively described via the transient electron temperature
and density in gold considering both intraband and interband
transitions at the surface plasmon resonance. This novel microscopy
technique enables background-free detection of the complex
susceptibility change even in highly scattering environments and can
be readily applied to any metal nanostructure.
\end{abstract}


\maketitle

Metallic nanoparticles exhibit morphology-dependent electromagnetic
resonances also called surface plasmon resonances (SPR) which couple
to propagating light. These resonances originate from a coherent
oscillation of electrons in the metal where the restoring force is
due to the electric field created by the corresponding charge
displacement. Since this electric field depends on the morphology,
the SPR frequency and linewidth is a function of the shape and size
of the NP and its dielectric
environment\,\cite{JainJPCB06,BerciaudNL05,MuskensPRB08}. The
resulting local optical resonances can be exploited to image
metallic NPs with high spatial resolution and to probe nanoscale
regions in the NP vicinity through the local electric field of the
resonance. Especially gold NPs are ideal optical labels for
biological applications owing to their bio-compatibility and
photostability, and much effort has been devoted recently to develop
techniques capable of detecting gold NPs in cells and tissues with
high contrast and sensitivity to the single particle
level\,\cite{BoyerScience02,DijkPCCP06,LindforsPRL04,LippitzNL05,ScwartzNL09,MasiaOL09}.

Ultrafast optical spectroscopy of metallic NPs is an intriguing area
of research investigating the correlated electronic and vibrational
dynamics in nanosized metals. Due to limited detection sensitivity,
most of the experiments performed so
far\,\cite{VoisinPRB04,HartlandChemRev11} used large ensembles of
metallic NPs, with the associated difficulty of inhomogeneous NP
size, shape and environment fluctuations. Only few reports on
ultrafast pump-probe spectroscopy on single small metallic
NPs\,\cite{MuskensNL06,VanDijkPRL05,VanDijkOE07,HartlandChemRev11}
are available to date. Importantly, these experiments lacked the
ability to extract the change in the NP dielectric function as a
{\it complex} quantity separating its real and imaginary
part\,\cite{noteGNP}. This however is the key physical quantity of
interest, being a function of the electron and lattice temperatures
transiently changing during the thermalization
dynamics\,\cite{VoisinPRB04}.

\begin{figure*}
\includegraphics*[]{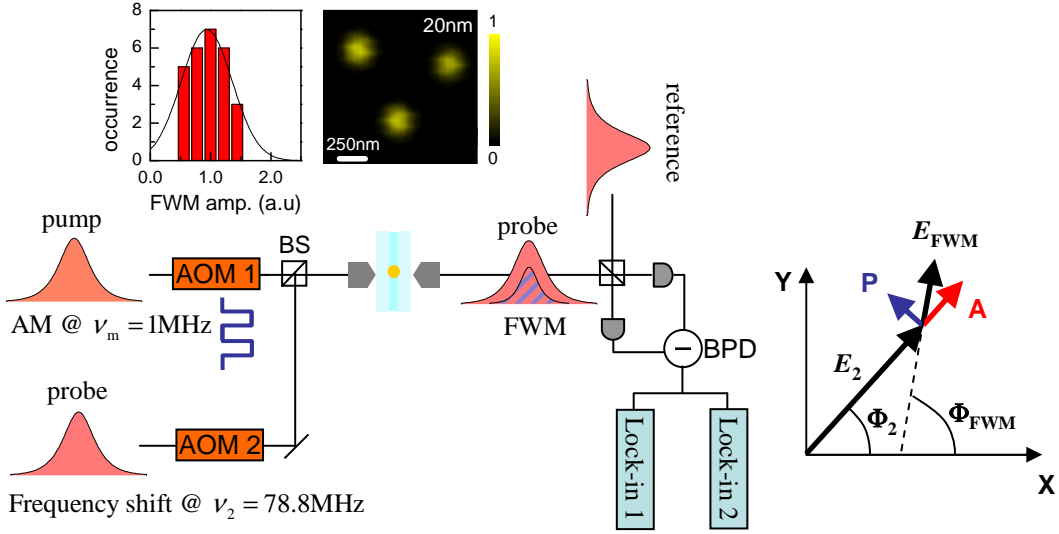}
\caption{Sketch of the phase-resolved transient four-wave mixing
micro-spectroscopy technique. A pump pulse train excites the gold NP
SPR with a temporally modulated intensity. The resulting change in
the NP optical properties is probed by a second pulse train
frequency shifted by an acousto-optic modulator (AOM2) and the FWM
field is collected in transmission and detected interferometrically
at the corresponding mixed frequency using an unshifted reference.
BS: 50:50 beam splitter, BPD: balanced photodiodes. Top inset: FWM
imaging of 20\,nm gold NPs resonantly excited and probed at 550\,nm
and corresponding histogram with a Gaussian fit showing a monomodal
size distribution (the deduced relative size distribution is 30\%).
Right: Sketch of the phase-resolved detection using two dual-channel
lock-in amplifiers to measure the $X$ and $Y$ components of the
probe field $E_2$ and FWM field $E_{\rm FWM}$. \label{fig1}}
\end{figure*}

To overcome these limitations we have developed a phase-sensitive
transient four-wave mixing microscopy technique as sketched in
Fig.\,\ref{fig1}. Laser pulses of $\sim150$\,fs Fourier-limited
duration with a tuneable center wavelength from 540\,nm to 590\,nm
and $\nu_{\rm L}=76.1$\,MHz repetition rate are split into three
beams. One beam acts as pump and excites the NP with an intensity
which is temporally modulated by an acousto-optic modulator (AOM)
driven with an amplitude modulation of frequency $\nu_{\rm
m}=1$\,MHz. This implementation replaces the two beam interference
used in our previous work\,\cite{MasiaOL09}, providing a stable
phase of the modulation. The change in the NP optical properties
induced by this excitation are probed by a second pulse at an
adjustable delay time $\tau$ after the pump pulse. A FWM signal
proportional to $E_{1}E_{1}^{*}E_{2}$ with $E_{1}$, $E_{2}$ electric
fields of the pump and probe, respectively, is collected in
transmission and detected interferometrically using the third beam
acting as reference. We used a heterodyne scheme to discriminate the
FWM field from pump and probe pulses and to detect amplitude and
phase of the field. By up-shifting the probe optical frequency via a
second AOM (driven with a constant amplitude at a frequency of
$\nu_{2}=78.8$\,MHz) we detected the interference of the FWM with
the reference at the frequency $\nu_{2}-\nu_{\rm m}-\nu_{\rm
L}=1.7$\,MHz by a dual-channel lock-in amplifier. We simultaneously
detect the interference of the transmitted probe with the reference
at $\nu_{2}-\nu_{\rm L}=2.7$\,MHz using a second dual-channel
lock-in, and deduced the amplitude ($A$) and phase ($P$) components
of the FWM field relative to the transmitted probe (see sketch in
Fig.\,\ref{fig1} and supplementary material (SM)). We have
previously demonstrated background-free FWM imaging in a highly
scattering and fluorescing environment, and sensitivity to the
single NP level with a spatial resolution significantly surpassing
one photon diffraction\,\cite{MasiaOL09}. An example of FWM imaging
of gold NPs of 20\,nm diameter resonantly excited and probed at
550\,nm is shown in the inset of Fig.\,\ref{fig1} together with an
histogram indicating a monomodal size distribution well separated
from the noise level (here $\sim1/100$ of the NP signal), confirming
that each spot corresponds to a single NP.

To explain how the detection of $A$ and $P$ enables us to
distinguish the transient changes of the real and imaginary part of
the dielectric function of a single NP, we write the induced
polarization field at the particle as $p=\epsilon_{0}\epsilon_{\rm
d}{\alpha}E_{2}$ where $\epsilon_{0}$ is the vacuum permittivity,
$\epsilon_{\rm d}$ is the dielectric constant of the medium
surrounding the NP and $\alpha$ is the particle polarizability. For
a particle radius $R$ much smaller than the wavelength of light
(Rayleigh limit), the polarizability is given by
$\alpha=4{\pi}R^3\frac{\epsilon-\epsilon_{\rm
d}}{\epsilon+2\epsilon_{\rm d}}$ with the particle dielectric
constant $\epsilon$. We denote $\epsilon_{\rm
d}\frac{\epsilon-\epsilon_{\rm d}}{\epsilon+2\epsilon_{\rm
d}}=\widetilde{\epsilon}$ as an effective dielectric constant, and
the FWM field in terms of the nonlinear polarization from the
pump-induced change $\Delta\widetilde{\epsilon}$, ie $E_{\rm
FWM}\propto\Delta\widetilde{\epsilon}E_{2}$. The complex dielectric
constant is then expressed in terms of its amplitude and phase
$\Delta\widetilde{\epsilon}=|\Delta\widetilde{\epsilon}|\exp(i\varphi)=\Delta\widetilde{\epsilon_{\rm
R}}+i\Delta\widetilde{\epsilon_{\rm I}}$, as well as the probe field
$E_2=|E_2|\exp(i\phi)$, such that $E_{\rm
FWM}\propto|\Delta\widetilde{\epsilon}||E_2|\exp(i(\phi+\varphi))$.
Hence $\varphi$ is the phase difference between the FWM field and
the probe field at the particle position. The transmitted probe in
the far field acquires a phase shift compared to a spherical wave of
the FWM field, known as Gouy phase shift for a Gaussian beam. The
Gouy phase shift is $-\pi/2$ from the focus to infinity, hence in
the far-field the FWM field is phase-shifted relative to the
transmitted probe by $\varphi+\pi/2$. The amplitude modulation of
the transmitted probe is then (see sketch in Fig.\,\ref{fig1})
${A}\propto|\Delta\widetilde{\epsilon}|\cos(\varphi+\pi/2)=-|\Delta\widetilde{\epsilon}|\sin(\varphi)=-\Delta\widetilde{\epsilon_{\rm
I}}$ and the phase modulation is
${P}\propto|\Delta\widetilde{\epsilon}|\sin(\varphi+\pi/2)=|\Delta\widetilde{\epsilon}|\cos(\varphi)=\Delta\widetilde{\epsilon_{\rm
R}}$. Note that the particle absorption cross-section is
proportional to the imaginary part of $\alpha$ as can be deduced
from Mie theory in the small particle limit, hence $A$ is a measure
of the change in absorption cross-section.

\begin{figure}
\includegraphics*[width=7cm]{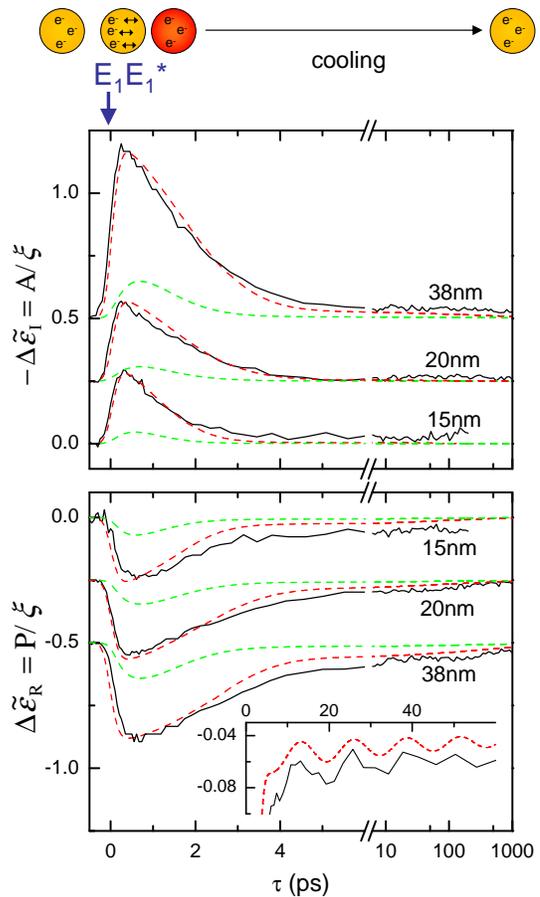}
\caption{Transient changes of the real
($\Delta\widetilde{\epsilon_{\rm R}}$) and imaginary
($\Delta\widetilde{\epsilon_{\rm I}}$) part of the dielectric
function of a single gold NP resonantly excited and probed at
550\,nm, for different NP diameters as indicated. Pump (probe)
fluence is 0.60\,J/m$^{2}$ (0.11\,J/m$^{2}$) for the 15\,nm and
20\,nm NPs and 0.65\,J/m$^{2}$ (0.05\,J/m$^{2}$) for the 38\,nm
particle. Acquisition time per point is 200\,ms. Dashed lines are
corresponding calculations (see text), with the green lines
neglecting the heating by interband absorption. Curves are
vertically displaced for clarity. The top sketch illustrates the SPR
excitation by the pump ($E_1$) and the subsequent heating and
cooling dynamics monitored by the probe. The inset shows coherent
phonon oscillations for the 38\,nm NP.\label{fig2}}
\end{figure}

The measured $A$ and $P$ versus pump-probe delay for single gold NPs
of various sizes resonantly excited and probed at 550\,nm are shown
in Fig.\,\ref{fig2}. Both exhibit an initial dynamics in the
picosecond time scale followed by a decay in the hundreds of
picoseconds. Qualitatively, one can understand these dynamics as
being related to the pump-induced increase in the electron
temperature which provokes a broadening and a shift of the SPR
probed by $E_2$. The picosecond decay is related to the
thermalization of the hot electron gas with the cooler lattice by
electron-phonon coupling, while the thermalization of the heated NP
with the cooler surrounding is providing the subsequent decay. To
achieve a quantitative description of the measured dynamics we
modeled $\widetilde{\epsilon}$ using the gold NP dielectric function
\begin{equation}\label{epsilon}
    \epsilon=1-\frac{\omega_{\rm
p}^2}{\omega(\omega+i\Gamma)}+\epsilon^{\rm b}(\omega)
\end{equation}
where $\omega_{\rm p}^2=n_{\rm e}e^2/\epsilon_0m_0$ is the plasma frequency ($n_{\rm e}$, $e$ and $m_0$ being the
conduction electron density, charge and effective mass respectively), $\Gamma$ is the inverse of the electron
relaxation time, and $\epsilon^{\rm b}$ is the contribution due to bound electrons associated with interband
transitions from the d-bands to the conduction band. At room temperature in the absence of optical excitation $\Gamma$
can be approximated as\,\cite{VoisinJPCB01}
\begin{equation}\label{Gamma}
\Gamma_0=\gamma_0+g\frac{v_{\rm F}}{R},
\end{equation}
where $\gamma_0$ is the bulk damping rate, $v_{\rm F}=1.4\times10^{6}$m/s is the Fermi velocity, and $g$ parametrizes
surface damping effects. For bulk gold we used $\hbar\omega_{\rm p}=9$\,eV and $\hbar\gamma_0=70$\,meV, well
reproducing the Drude part of the bulk dielectric constant measured by Johnson and Christy\,\cite{JohnsonPRB72}. These
values also reproduce experimental absorption spectra of single gold nanoparticles in the
literature\,\cite{BerciaudNL05}. The parameter $g$ varies in the literature\cite{BerciaudNL05,MuskensPRB08} from 0.2 to
2.2, and even between individual NPs within the same experiment. This variance is likely to be related to the atomistic
NP surface structure/faceting\,\cite{MuskensPRB08} hence we use $g$ as free parameter falling in the range of
previously reported values. In the presence of optical excitation we then model the change in the electron temperature
$T_{\rm e}$ and lattice temperature $T_{\rm l}$ using a two-temperature model\,\cite{VoisinJPCB01,MuskensNL06}
\begin{equation}\label{Tel}
C_{\rm e}(T_{\rm e})\frac{dT_{\rm e}}{dt}=-G_{\rm ep}(T_{\rm e}-T_{\rm l})+P(t),
\end{equation}
\begin{equation}\label{Tlat}
C_{\rm l}\frac{dT_{\rm l}}{dt}=G_{\rm ep}(T_{\rm e}-T_{\rm l})-\dot{Q},
\end{equation}
where $C_{\rm e}$ is the electronic heat capacity, well described by $C_{\rm e}={\gamma}T_{\rm e}$ for \cite{LinPRB08}
$T_{\rm e} < 3000$\,K, $C_{\rm l}$ is the lattice heat capacity, and $G_{\rm ep}$ is the electron-phonon coupling
constant. $C_{\rm e}$ and $C_{\rm l}$ are taken as the bulk value scaled to the NP volume (see SM). $\dot{Q}$ describes
the heat loss from particle to the surrounding and is calculated using the diffusion equation for the temperature of
the surrounding medium (see SM). $P(t)$ describes the excitation process which has two components, $P_{\rm f}$ and
$P_{\rm b}$, related to free and bound electrons, respectively. The bound electron contribution has to be considered
since in spherical gold NPs the onset of interband transitions from d-bands to the conduction band at the Fermi surface
is energetically close to the SPR.\cite{VoisinJPCB01} $P_{\rm f}$ is modeled as a rapid ($\sim10$\,fs, quasi
instantaneous for the 150\,fs pulses used) transfer of energy from the resonantly excited SPR to single electron
excitations of the conduction band, which thermalize with the free-electron gas creating a hot Fermi-Dirac distribution
on a characteristic time scale $\tau_{\rm f}$\,\cite{VoisinJPCB01,VoisinPRB04}. Similarly $P_{\rm b}$ originates from
the transfer of energy from the SPR to interband excitations and consequent heating of the free electron gas via Auger
electron-hole recombination with a characteristic time $\tau_{\rm b}$ (see SM). Noticeably, most pump-probe experiments
in previous literature use a pump photon energy below the onset of interband transitions, and thus neglect the latter
effect\,\cite{VoisinPRB04,HartlandChemRev11}. The time integral of $P(t)$ is equal to the absorbed energy calculated
from the NP absorption cross-section and the pump pulse fluence in the experiment.

The non-equilibrium electron and lattice temperatures modify the electron-electron and electron-phonon scattering
processes, which can be described\cite{SmithPRB82} by $\Gamma=\Gamma_0+\gamma_{\rm ep}+\gamma_{\rm ee}$ with
\begin{equation}\label{Gammaee}
\gamma_{\rm ep}(T_{\rm l})=b_{\rm ep}\hbar^2\omega^2(T_{\rm l}-T_0),
\end{equation}
and
\begin{equation}\label{Gammaep}
\gamma_{\rm ee}(T_{\rm e})=b_{\rm ee}\hbar^2\omega^2(T_{\rm e}^2-T_0^2).
\end{equation}
using the equilibrium temperature $T_0$. The electron and lattice temperatures also affect $\epsilon^b$. To estimate
this effect we calculated the imaginary part $\epsilon_{2}^b$ of $\epsilon^b$ due to interband transitions near the X
and L points, starting from the description in Refs.\,\onlinecite{GuerresiPRB75,RoseiPRB74}. The temperature dependence
of $\epsilon^b$ results from the Fermi distribution and a spectral broadening proportional to $\Gamma$ (see SM). The
real part of $\epsilon^b$ is calculated from $\epsilon_{2}^b$ using Kramers-Kronig.

The transient change in $\epsilon$ modifies $\widetilde{\epsilon}$ and is the source of the measured
$\Delta\widetilde{\epsilon_{\rm I}}$ and $\Delta\widetilde{\epsilon_{\rm R}}$ via $A=-\xi\Delta\widetilde{\epsilon_{\rm
I}}$ and $P=\xi\Delta\widetilde{\epsilon_{\rm R}}$ (see Fig.\,\ref{fig2}) with a constant $\xi$\,\cite{noteGNP2}. We
are able to consistently model all measured dynamics for particle sizes up to 40\,nm diameter and for different pump
excitation intensities in the $0.16-1.3$\,J/m$^{2}$ range with only few particle-dependent ($g$, $G_{\rm ep}$) and
particle independent ($b_{\rm ee}$, $b_{\rm ep}$, $\tau_{\rm f}$, $\tau_{\rm b}$) parameters. The gold band--structure
parameters to calculate $\epsilon^b$ are taken from literature\,\cite{GuerresiPRB75,RoseiPRB74}. The proportionality
term $\xi$ scales with $R^3$ and is independent of the pump intensity as expected from the particle polarizability
$\alpha$. Larger particles have a polarizability which deviates significantly from the Rayleigh limit used here, and we
found that FWM dynamics measured on 100\,nm diameter particles can no longer be described by our model (see SM). We
calibrated the size of the 38\,nm NP investigated in Fig.\,\ref{fig2} using the period of the coherent phonon
oscillations observed in $\Delta\widetilde{\epsilon_{\rm R}}$ (see inset in Fig.\,\ref{fig2}). These are due to the
modulation of the plasma frequency by the breathing vibrational mode of the nanoparticle with an oscillation period
proportional to the particle radius\,\cite{VoisinJPCB01,VanDijkPRL05,HartlandChemRev11}(see SM). On NPs smaller than
20\,nm the oscillation period was not resolved, possibly because its time scale becomes comparable to the lattice
expansion which is no longer impulsively exciting the vibrational mode hence its amplitude falls below the measurement
noise. In this case the particle size was derived from the strength of the measured FWM field relative to that of the
38\,nm particle, consistently with $\xi \propto R^3$.

\begin{figure}
\includegraphics*[width=7cm]{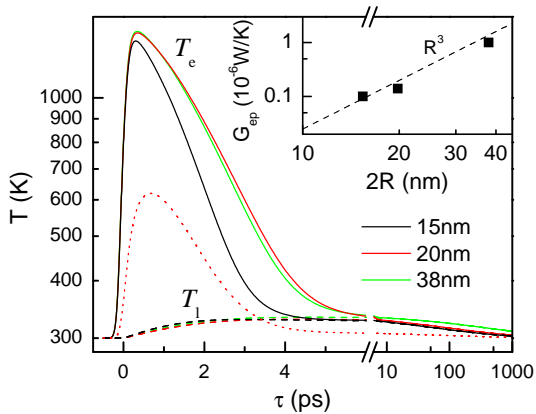}
\caption{Transients of the electron temperature (solid lines) and lattice temperature (dashed lines) for different NP
sizes as indicated, calculated using Eq.\,\ref{Tel},\,\ref{Tlat} for a pump fluence of 0.60\,J/m$^{2}$.  The dotted
curve is the electron temperature of a 20\,nm NP when the excitation term involving bound electrons is neglected. The
inset shows the electron--phonon coupling parameter used in the calculations. The dashed line in the inset indicates
the volumetric scaling.\label{fig3}}
\end{figure}

We use $\tau_{\rm f}=500$\,fs independent on particle size in the 15-40\,nm diameter range from
Ref.\,\onlinecite{VoisinJPCB01}, $b_{\rm ee}=0.89 \times10^{7}$\,s$^{-1}$eV$^{-2}$K$^{-2}$ and $b_{\rm ep}=0.5
\times10^{11}$\,s$^{-1}$eV$^{-2}$K$^{-1}$ consistent with Ref.\,\onlinecite{SmithPRB82}, and $g$ in the range\, 1.8-1.9
consistent with Ref.\,\onlinecite{MuskensPRB08}.

The fast rise of $\Delta\widetilde{\epsilon_{\rm I}}$ and $\Delta\widetilde{\epsilon_{\rm R}}$ observed in our data
reveals a fast d-state Auger recombination time of $\tau_{\rm b}=70$\,fs $\ll \tau_{\rm f}$. While this time has not
been measured previously in Au, it is within the energy-dependent range of 250-20\,fs measured in
Cu\,\cite{KnoeselPRB98}. To clarify the significance of interband transitions in the SPR resonant excitation, we show
in Fig.\,\ref{fig2} also simulations (green lines) neglecting the heating $P_{\rm b}$ in Eq.\,\ref{Tel} due to the
interband transitions. The effect of interband transitions in the excitation is important to account not only for the
magnitude of the measured change, but also its dynamics, particularly for the observed fast rise time. We could
simulate the measured rise time without the interband contribution only when using $\tau_{\rm f}<100$\,fs, much faster
than the well characterized 500\,fs electron thermalization dynamics\,\cite{VoisinJPCB01} in bulk gold and gold NPs
larger than 10\,nm. The faster Auger recombination of the d-holes thus gives rise to a qualitatively different initial
dynamics, and can be attributed to a higher density of states available in the scattering involving the lower-lying
d-states.

\begin{figure*}[]
\includegraphics*[width=12cm]{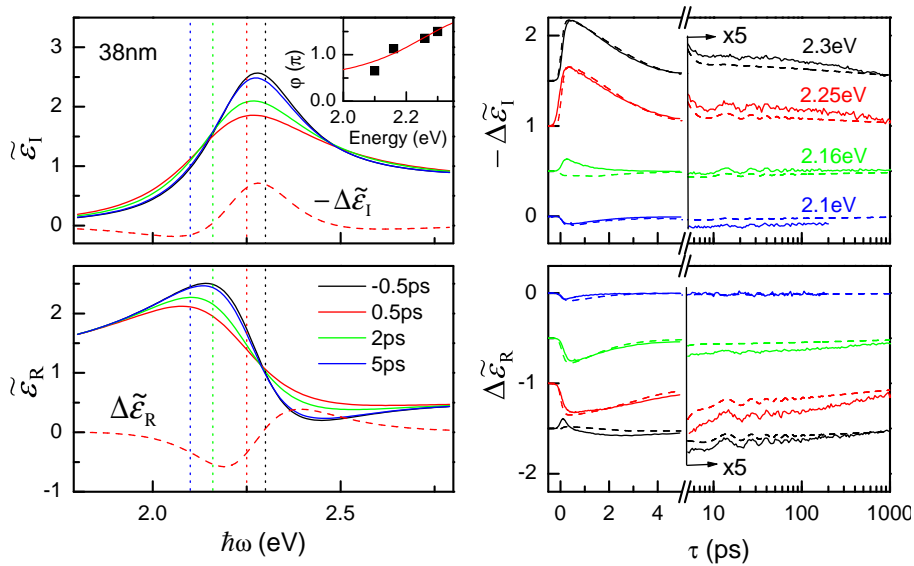}
\caption{Left: Spectrally-resolved real and imaginary part of the dielectric constant $\tilde\epsilon$ simulated for
the 38\,nm NP at various delay times, for an excitation at 550\,nm. Dashed lines are the differential spectra at
$\tau=0.5$\,ps. Vertical dotted lines indicate the excitation and detection photon energies used in the experiment on
the right. In the inset, the phase of $\tilde\epsilon$ is shown at $\tau=0.5$\,ps (symbols: experiment, line:
simulation). Right: Measurement and simulations of the transient changes of the real and imaginary part of the
dielectric function at different photon energies, as indicated. The pump fluence was 0.48\,J/m$^{2}$ at 540\,nm and
0.6\,J/m$^{2}$ for all other wavelengths. Curves are vertically displaced for clarity.\label{fig4} }
\end{figure*}

It is instructive to examine the temperature transients of $T_{\rm
e}$ and $T_{\rm l}$ plotted in Fig.\,\ref{fig3}. Noticeably the
electron temperature reaches a maximum value which is almost
independent of NP size. This is a consequence of the volumetric
dependence of both the electron heat capacity and the absorption
cross-section, balancing each other. The decay of $T_{\rm e}$ on the
picosecond time scale coincides with the rise of $T_{\rm l}$ due to
electron-phonon coupling, and the electron temperature after
thermalization is much lower due to the significantly larger lattice
heat capacity ($C_{\rm l}\gg C_{\rm e}$). We find that $G_{\rm ep}$
also has a nearly volumetric dependence, and in turn the
electron-phonon dynamics exhibits only a weak size dependence in the
size range 15-40\,nm investigated here, consistent with reports in
the literature\,\cite{ArbouetPRL03}. We also observe that the
cooling with the environment is size dependent, with smaller
particles cooling faster\,\cite{HartlandChemRev11}. We point out
that the {\it average} temperature increase at the particle surface
is only 1.3\,K, 1.9\,K, and 4.7\,K for 15\,nm, 20\,nm and 38\,nm
diameters respectively and can be lowered to 1.3\,K for 38\,nm
diameter at the lowest ($0.16$\,J/m$^{2}$) pump excitation used in
the experiment. Importantly, such low average photothermal heating
indicates that our technique is compatible with live cell
applications. Finally, the effect of the interband excitation
$P_{\rm b}$ is shown by plotting $T_{\rm e}$ in its absence (dotted
curve in Fig.\,\ref{fig3}), which shows a delayed rise and reaches
only about half of the maximum electron temperature.

To visualize the origin of the FWM signal in terms of broadening and shift of the SPR we plot in Fig.\,\ref{fig4} the
spectra of $\widetilde{\epsilon_{\rm I}}$ and $\widetilde{\epsilon_{\rm R}}$ at different delay times for an excitation
at 550\,nm. The negative change of $\widetilde{\epsilon_{\rm I}}$ probed at 550\,nm (see dotted red vertical line in
Fig.\,\ref{fig4}) is mainly a result of the broadening of the SPR and in turn quenching of the resonance. Conversely,
the negative change of $\widetilde{\epsilon_{\rm R}}$ at 550\,nm is a combination of the SPR broadening and its
frequency shift toward lower energies. Measurements and simulations of $\Delta\widetilde{\epsilon_{\rm I}}$ and
$\Delta\widetilde{\epsilon_{\rm R}}$ excited and probed at 540\,nm ($\hbar\omega=2.3$\,eV) and 575\,nm
($\hbar\omega=2.16$\,eV) elucidate the cases of $\Delta\widetilde{\epsilon_{\rm I}}$ being mostly sensitive to the SPR
broadening and $\Delta\widetilde{\epsilon_{\rm R}}$ to the SPR shift, and vice versa. Specifically we see that at
2.3\,eV, $\Delta\widetilde{\epsilon_{\rm R}}$ is a sensitive probe of the modulation of the SPR by the breathing
vibrational mode, while $\Delta\widetilde{\epsilon_{\rm I}}$ is essentially unaffected by it. We also performed
measurements and simulations at 590\,nm ($\hbar\omega=2.10$\,eV) where $\Delta\widetilde{\epsilon_{\rm I}}$ changes
sign. Noticeably, the initial dynamics measured at different wavelengths is consistently reproduced by our model of the
fast Auger recombination of the d-holes. Finally, to illustrate that the novel imaging technique presented here can be
used to sense SPR shifts relative to the probe wavelength in an intrinsic ratiometric way independent of the signal
strength, we show in the inset of Fig.\,\ref{fig4} the phase $\varphi$ of
$\Delta\widetilde{\epsilon}=|\Delta\widetilde{\epsilon}|\exp(i\varphi)$ from the ratio between its imaginary and real
part at $\tau=0.5$\,ps. $\varphi$ changes by nearly 180 degrees over only 50\,nm wavelength as a direct manifestation
of the transition from $\Delta\widetilde{\epsilon_{\rm R}}=0$, $\Delta\widetilde{\epsilon_{\rm I}}<0$
($\varphi=3\pi/2$, at 540\,nm) to $\Delta\widetilde{\epsilon_{\rm I}}=0$, $\Delta\widetilde{\epsilon_{\rm R}}<0$
($\varphi=\pi$, at 575\,nm), to $\Delta\widetilde{\epsilon_{\rm I}}>0$, $\Delta\widetilde{\epsilon_{\rm R}}<0$
($\pi/2<\varphi<\pi$, at 590\,nm).

Our technique is applicable to any NP shape to reveal new physics insights in the electron and lattice dynamics of
novel metallic nanostructures. Importantly it is compatible with living cell applications and offers background--free
time-resolved detection of the full complex change of the dielectric constant. The intrinsic ratiometric information in
the signal phase enables sensing applications such as monitoring nanoscale distance changes with plasmon
rulers\,\cite{SonnichsenNatBiot05} in highly scattering environments such as cells and tissues.

\section{Methods}

Commercially available gold NPs in aqueous suspension (BB
International) were drop cast on a glass cover slip at sufficiently
low density so that regions with well separated individual NPs were
formed, and subsequently embedded in a mounting medium (Cargille
Meltmount) index-matched to the oil-immersion microscope objectives
used for high-resolution imaging (see also
Ref.\,\onlinecite{MasiaOL09}). The sample is scanned using an xyz
piezoelectric stage (Physik Instrumente Nanocube).

Optical pulses were provided by the intra-cavity frequency-doubled optical parametric oscillator (APE-PP2) pumped by a
Ti:Sapphire oscillator (Coherent Mira). Pulses were linearly polarized vertically (V) or horizontally (H) in the
laboratory system using $\lambda/4$ and $\lambda/2$ waveplates, and a VHH polarization configuration for $E_{1}$,
$E_{2}$ and the reference field was used to suppress background from detector nonlinearities\,\cite{MasiaOL09}.

\begin{acknowledgments}
F.M. acknowledges financial support from the European Union (Marie
Curie grant agreement PIEF-GA-2008-220901) and the Welcome Trust
(VIP award). P.B. acknowledges the EPSRC UK Research Council for her
Leadership fellowship award (grant n. EP/I005072/1).
\end{acknowledgments}


\begin{thebibliography}{25}
\bibitem [{not(){\natexlab{b}}}]{noteGNP2}%
  \BibitemOpen
  \bibinfo {note} {To represent the
  measurement conditions in the detection, the calculated
  $\widetilde{\epsilon}$ is convoluted with the pulse intensity gaussian
  profile.}\BibitemShut {Stop}%
\end{thebibliography}

%

\end{document}



\title{Supplementary Material}




\author{Francesco Masia}
\affiliation{Cardiff University School of Physics and Astronomy, The
Parade, Cardiff CF24 3AA, United Kingdom}\affiliation{Cardiff
University School of Biosciences, Museum Avenue, Cardiff CF10 3AX,
United Kingdom}
\author{Wolfgang Langbein}
\affiliation{Cardiff University School of Physics and Astronomy, The
Parade, Cardiff CF24 3AA, United Kingdom}
\author{Paola Borri}
\email{borrip@cardiff.ac.uk}\affiliation{Cardiff University School
of Biosciences, Museum Avenue, Cardiff CF10 3AX, United
Kingdom}\affiliation{Cardiff University School of Physics and
Astronomy, The Parade, Cardiff CF24 3AA, United Kingdom}



\maketitle

\section{Setup description}
Optical pulses are provided by the intra-cavity frequency doubled
signal of an optical parametric oscillator (APE-PP2) pumped by a
femtosecond Ti:Sapphire oscillator (Coherent Mira) of $\nu_{\rm
L}=76.1$\,MHz repetition rate. We used pulses with a tuneable center
wavelength from 540\,nm to 590\,nm and a Fourier-limited duration
ranging from 150\,fs to 180\,fs (intensity full-width at
half-maximum) depending on the wavelength (corresponding to 2.5\,nm
spectral width at 550\,nm). The pulse train at the output of the
optical parametric oscillator is split into three beams having all
the same center wavelength (degenerate FWM scheme). The delay times
between the pulses are controlled with 10\,fs accuracy by linear
stages. Pump and probe pulses of fields $E_1$ and $E_2$ respectively
are recombined into the same spatial mode and focused onto the
sample by an oil-immersion microscope objective (MO) of 1.25
numerical aperture (NA) to achieve high spatial resolution. The
sample is positioned with respect to the focal volume of the
objective by a $xyz$ piezoelectric stage with 10\,nm resolution
(Physik Instrumente Nanocube). The FWM signal is collected in
transmission by a second MO opposite to the sample, recombined in a
beamsplitter with a reference pulse of adjustable delay, and the
resulting interference is detected by balanced silicon photodiodes
and a lock-in amplifier (SRS SR844). Discrimination of the FWM
signal is achieved using a heterodyne scheme. In this scheme two
acousto-optic modulators (AOMs) are used to upshift the optical
frequencies of $E_{1,2}$ by the radiofrequencies $\nu_1=80$\,MHz and
$\nu_2=78.8$\,MHz, respectively. The intensity of the first pulse is
modulated with close to unity depth by applying a $\nu_{\rm
m}=1$\,MHz square wave amplitude modulation to the first AOM. The
interference of the FWM signal with the unshifted reference field
gives rise to a beat note at the radiofrequency $\nu_2-\nu_{\rm m}$
and its sidebands separated by the repetition rate $\nu_{\rm L}$ of
the pulse train. Using an electronically synchronized reference
created by mixing the AOM drive frequencies with the laser
repetition rate, we detect the interference of the FWM with the
reference by a dual-channel lock-in amplifier at $\nu_{2}-\nu_{\rm
m}-\nu_{\rm L}=1.7$\,MHz. Simultaneously, we detect the interference
of the $E_{2}$ field with the reference using a second lock-in
amplifier at $\nu_{2}-\nu_{\rm L}$. We employed a cross-polarization
scheme to reject background originating from nonlinearities of the
photodiodes, in which the polarization state of $E_{1}$ was adjusted
by a combination of $\lambda/4$ and $\lambda/2$ waveplates to be
cross-linearly polarized to $E_2$ and the reference beam. Polarizers
in front of the photodiodes transmit the FWM, $E_2$, and reference
fields, but block the pump field $E_{1}$.

\section{Amplitude and phase parts of the
FWM field} Two dual-channel lock-in amplifiers are used to measure
the $X$ and $Y$ components of $E_2$ (probe field) and $E_{\rm FWM}$
(FWM field), relative to the reference field, as shown in the paper
in Fig.\,1. They are electronically referenced to the frequencies
$\nu_{2}-\nu_{\rm L}$ and $\nu_{2}-\nu_{\rm m}-\nu_{\rm L}$,
respectively, in a phase-stable manner. We define the amplitude part
A as the projection of $E_{\rm FWM}$ onto $E_2$ and the phase part P
as its orthogonal component:

\begin{eqnarray}\label{AMdef}
 \nonumber 
  A &=& \frac{X_{2}X_{\rm FWM}+Y_{2}Y_{\rm FWM}}{|E_2|^2} \\
  P &=& \frac{X_{2}Y_{\rm FWM}-Y_{2}X_{\rm FWM}}{|E_2|^2}
\end{eqnarray}

We calibrate the relative reference phase of the two lock-in amplifiers using a known amplitude and phase modulation of
the probe. For this purpose we measure an electrically pumped quantum-dot optical amplifier, which shows pump-induced
changes of the gain and refractive index\,\cite{CesariJQE09} resulting in large changes of amplitude ($>10\%$) and
phase ($>10^\circ$) of the transmitted probe field. These changes can be measured with the FWM experiment, but also
directly by measuring $X_2$ and $Y_2$ with and without the pump using a slow (5\,Hz) modulation. Both measurements have
to yield the same result, since the response of the optical amplifier is the same at 1\,MHz and 5\,Hz. This requirement
determines the phase offset $\phi_{\rm o}$ between the two lock-in amplifiers in the FWM experiment, which is taken
into account by transforming the values $X_2^{(M)},Y_2^{(M)}$ directly measured by the lock-in into
$X_2=X_2^{(M)}\cos(\phi_{\rm o})-Y_2^{(M)}\sin(\phi_{\rm o}), Y_2=Y_2^{(M)}\cos(\phi_{\rm o})+X_2^{(M)}\sin(\phi_{\rm
o})$.  The phase offset reproducibility during the course of the experiments was found to be a few degrees
(corresponding to a few nanoseconds delay), as expected from the RF electronics used to create the lock-in reference
signals.

\section{Model of the response}

\subsection{Excitation}\label{excitation}

The response of the GNP polarizability to an optical excitation is dominated by the dynamics of the electron
distribution. The optical excitation creates an internal optical electric field in the material according to the
morphology and dielectric material response. This field is absorbed creating non-thermal electron distributions, which
thermalize by electron-electron scattering with characteristic thermalization times, establishing a hot Fermi-Dirac
distribution with broadened Fermi-edge \cite{VoisinJPCB01,VoisinPRB04}.

We can distinguish two different electronic excitations creating a heating power $P$ for the electron distribution,
related to free ($P_{\rm f}$) and bound ($P_{\rm b}$) electrons. The latter is important in our experiments since the
onset of interband transitions from d-bands to the conduction band at the Fermi surface is close to the SPR in
spherical gold NPs\,\cite{VoisinJPCB01}. We emphasize that the dynamics from a single particle excitation to a
thermalized electron distribution by cascaded electron-electron scattering processes is non-trivial, and depends on the
electronic band-structure and the initial single particle excitation. It is thus expected that different initial single
particle excitations are described by different effective thermalization times $\tau$.

The total energy absorbed by the NP is given by $I\sigma_{\rm abs}$,
where $I$ is the pump pulse fluence and $\sigma_{\rm abs}$ is the NP
absorption cross section. It is distributed among the different
single particle excitations ($\rm SPE$) related to the free electron
gas ($\rm SPE = f$) and the interband transitions involving the X
and L point ($\rm SPE = X, L$) respectively. The fraction absorbed
by the corresponding SPE is given by the relative contribution
$f_{\rm SPE}$ to the imaginary part of the bulk dielectric constant
of the NP $\epsilon_2(\omega_{\rm exc})$ at the excitation frequency
$\omega_{\rm exc}$,
\begin{equation}\label{fractionholes}
f_{\rm SPE}=\frac{\epsilon^{\rm SPE}_{2}(\omega_{\rm exc})}{\epsilon_{2}(\omega_{\rm exc})}
\end{equation}
where $\epsilon^{\rm SPE}_{2}$ is the contribution in $\epsilon_2$ related to the SPE.

The SPR lifetimes are typically in the $\sim10$\,fs range and can be neglected compared to the pulse duration used in
our experiments. Assuming a Gaussian intensity profile of the pump pulse
\begin{equation}\label{Excitationfree}
I(t)\propto\frac{1}{\sqrt{\pi\sigma^2}}\exp\left(-t^2/\sigma^2\right),
\end{equation}
and a subsequent thermalization described as a Markovian process with time constant $\tau$, we find the heating powers
\begin{equation}\label{Power}
P_{\rm SPE}(t)=\frac{I\sigma_{\rm abs}f_{\rm SPE}}{2\tau_{\rm SPE}}\left[1+{\rm erf}\left(\frac{2t\tau_{\rm
SPE}-\sigma^2}{2\tau_{\rm SPE}\sigma}\right)\right]\exp{\left(\frac{\sigma^2-4t\tau_{\rm SPE}}{4\tau_{\rm
SPE}^2}\right)}
\end{equation}
where $\sigma$ is given by the pulse duration in the experiment measured for every wavelength (eg $\sigma=0.11$\,ps for
an intensity full-width at half-maximum of 0.18\,ps).

$P_{\rm f}$ originates from the intraband absorption of the free electron gas and has been described in the literature
extensively, having a thermalization time $\tau_{\rm f}\approx 500\,$fs.

$P_{\rm b}=P_{\rm X}+P_{\rm L}$ instead originates from the absorption of the SPR by interband transitions and
consequent Auger electron-hole recombination and heating of the free electron gas, which we model with a separate
thermalization time $\tau_{\rm b}=\tau_{\rm L}=\tau_{\rm X}$.

To calculate the transient hole densities which influence the dielectric response, we express
\begin{equation}\label{Excitationbound}
P_{\rm b}(t)=\left(n_{\rm X}^{\rm nth}(t)+n_{\rm L}^{\rm nth}(t)\right)\frac{V\hbar\omega_{\rm exc}}{\tau_{\rm b}}
\end{equation}
in terms of the densities of photoexcited holes $n_{\rm X}^{\rm nth},n_{\rm L}^{\rm nth}$ near the X and L points,
respectively, which lead to Auger recombination heating of the electron distribution. Here $V$ is the particle volume.
The densities evolve according to
\begin{equation}\label{holes}
\frac{dn_{\rm X,L}^{\rm nth}}{dt}=\frac{n_{\rm
X,L}}{\sqrt{\pi\sigma^2}}\exp{\left(-\frac{t^2}{\sigma^2}\right)}-\frac{n_{\rm X,L}^{\rm nth}}{\tau_{\rm b}}
\end{equation}
which represents the generation by the Gaussian pulse excitation and the decay due to Auger recombination. The
photogenerated densities $n_{\rm X,L}$ are given by
\begin{equation}\label{amplitudeholes}
n_{\rm X,L}=\frac{I\sigma_{\rm abs}f_{\rm X,L}}{V\hbar\omega_{\rm exc}}.
\end{equation}
At the resonant excitation wavelength of 550\,nm in our experiment we find (see section\,\ref{subsec:interband})
$f_{\rm X}=0.59$ and $f_{\rm L}=0.16$, i.e. 75\% of the excitation is absorbed by interband processes. This corresponds
for a 38\,nm diameter NP with a pump fluence of $I=0.32$\,J/m$^{2}$ to photogenerated hole densities of $n_{\rm
X}=3\times10^{-3}n_{\rm e}$, $n_{\rm L}=8\times10^{-4}n_{\rm e}$, with the free electron density $n_{\rm
e}=5.9\times10^{28}$\,m$^{-3}$. For the other excitation wavelengths used we find $f_{\rm X}=0.57$ and $f_{\rm L}=0.24$
at 540\,nm, $f_{\rm X}=0.59$ and $f_{\rm L}=0.09$ at 575\,nm, $f_{\rm X}=0.57$ and $f_{\rm L}=0.06$ at 590\,nm.

\subsection{Two-temperature model}

To describe the energy exchange between the optical excitation, electrons in the NP and lattice we use a
two-temperature model including heat diffusion with the surrounding medium\,\cite{HamanakaPRB01,RashidiJAP04}

\begin{equation}\label{Tel}
C_{\rm e}\frac{dT_{\rm e}}{dt}=-G_{\rm ep}(T_{\rm e}-T_{\rm l})+P_{\rm f}(t)+P_{\rm b}(t),
\end{equation}
\begin{equation}\label{Tlat}
C_{\rm l}\frac{dT_{\rm l}}{dt}=G_{\rm ep}(T_{\rm e}-T_{\rm l})+4{\pi}R^{2}\kappa_{\rm s}\left.\frac{{\partial}T_{\rm
s}}{{\partial}r}\right|_{r=R},
\end{equation}

\begin{equation}\label{Tsurr}
\frac{{\partial}T_{\rm s}}{{\partial}t}=D_{\rm
s}\frac{1}{r^2}\frac{\partial}{{\partial}r}\left[r^2\frac{{\partial}T_{\rm s}}{{\partial}r}\right],
\end{equation}
where $C_{\rm e}={\gamma}T_{\rm e}$ is the temperature dependent electronic heat capacity, $C_{\rm l}$ is the lattice
heat capacity, $G_{\rm ep}$ is the electron-phonon coupling constant. We use $\gamma/V=67.6$\,Jm$^{-3}$K$^{-2}$ and
$C_{\rm l}/V=2.49\times10^{6}$Jm$^{-3}$K$^{-1}$ according to literature\,\cite{HamanakaPRB01,RashidiJAP04}, and deduce
the particle values by multiplying these volumetric constants with the particle volume $V=\frac{4}{3}{\pi}R^{3}$.
$T_{\rm s}(r,t)$ is the temperature of the surrounding which obeys the radial heat diffusion Eqn.(\,\ref{Tsurr}),
$\kappa_{\rm s}=0.07\,$Wm$^{-1}$K$^{-1}$ is the thermal conductivity and $D_{\rm s}=0.1\times10^{-7}$m$^{2}$s$^{-1}$ is
the thermal diffusivity of the surrounding. Those values are obtained from the best fit to the FWM dynamics. The
thermal conductivity and diffusivity of the mounting medium surrounding the nanoparticles in our experiment (Cargille
Meltmount) are not known, but the fitted values are similar to the known values of PVC and epoxy resins. The initial
and boundary conditions are as follows: Electrons, lattice, and surrounding medium are initially at thermal
equilibrium, ie $T_{\rm e}(0)=T_{\rm l}(0)=T_{\rm s}(r,0)=T_0=300$\,K. The surrounding medium in contact with the
particle surface and the metal lattice are always at thermal equilibrium, $T_{\rm l}(t)=T_{\rm s}(R,t)$. The
surrounding medium far away from the particle is at ambient $T_{\rm s}(\infty,t)=300$\,K temperature.

The system of differential equations was solved in MATLAB using the function {\it pdepe}, which solves initial-boundary
value problems for parabolic-elliptic partial differential equations in one dimension.

\subsection{Contribution of interband transitions to the bulk dielectric constant }
\label{subsec:interband}

To calculate $\epsilon^{\rm b}_{2}$ we use the approach developed in Refs.\,\onlinecite{RoseiPRB74,GuerresiPRB75}. We
write $\epsilon^{\rm b}_{2}$ in terms of the joint density of states (Eqn.(9) in Ref.\,\onlinecite{GuerresiPRB75})

\begin{equation}\label{epsilon2}
\epsilon^{\rm b}_{2}=\frac{A_{\rm X}J_{\rm X}+A_{\rm L^{+}_{5+6}}J_{\rm L^{+}_{5+6}}+A_{\rm L^{+}_{4}}J_{\rm
L^{+}_{4}}}{{\hbar\omega}^2},
\end{equation}
where $A_{\rm X}$ is proportional to the square of the transition dipole matrix element $P_{\rm X}$ of the interband
transition from the d-band to the Fermi surface near the $X$ point of the band-structure. Similarly $A_{\rm
L^{+}_{5+6}}$ and $A_{\rm L^{+}_{4}}$ are proportional to the square of the transition dipole matrix elements near the
$L$ point, involving  the 5+6  and 4 d-bands respectively (see band structure calculated in
Ref.\,\onlinecite{ChristensenPRB71}). $J_{\rm X}$, $J_{\rm L^{+}_{5+6}}$, $J_{\rm L^{+}_{4}}$ are the corresponding
transition density of states which have a temperature dependence given by

\begin{equation}\label{jointDOS}
    J_{\rm i}=\int_{E^{\rm i}_{\rm min}}^{E^{\rm i}_{\rm
    max}}\frac{1}{k_{\rm i}(E)}(f_{\rm d}(E,T_{\rm e}) - f_{\rm p}(E,T_{\rm e}))dE
\end{equation}

with ${\rm i = X, L^{+}_{5+6}, L^{+}_{4}}$.

$f_{\rm p}$ and $f_{\rm d}$ are the occupation probabilities of the
p-band and d-band respectively

\begin{equation}\label{occupationp}
f_{\rm p}(E,T_{\rm e}) = \frac{1}{1+\exp{(\frac{E-{\Delta}E_{\rm
F}(T_{\rm e})}{k_{\rm B}T_{\rm
    e}}})}
\end{equation}
\begin{equation}\label{occupationd}
f_{\rm d}(E,T_{\rm e}) =
1-\frac{1}{1+\exp{(\frac{\hbar\omega-E+{\Delta}E_{\rm F}(T_{\rm
e})}{k_{\rm B}T_{\rm
    e}}})}
\end{equation}
and
\begin{equation}\label{kappaX}
k_{\rm X}(E)=\sqrt{\hbar\omega - E - \hbar\omega_7^X +
\frac{m_{p\parallel}^X}{m_{d\parallel}^X}(E-\hbar\omega_6^X)}
\end{equation}
\begin{equation}\label{kappaL}
k_{\rm L}(E)=\sqrt{\hbar\omega - E - \hbar\omega_{4^-}^L -
\hbar\omega_0+
\frac{m_{p\perp}^L}{m_{d\perp}^{L}}(E+\hbar\omega_{4^-}^L)}
\end{equation}
where $\hbar\omega$ is the photon energy, $\hbar\omega_0$ is $\hbar\omega_{5+6}^L$ or
$\hbar\omega_{4^+}^L$, and correspondingly $m_{d\perp}^{L}$ refers to either the 5+6 or the 4
d-bands. The Fermi level as a function of electron temperature $E_{\rm F}(T_{\rm e})$ is calculated
from the relationship

\begin{equation}\label{Fermi}
    N_{\rm e}=\int^{\infty}_{- \infty} \frac{{\rm DOS}(E)dE}{1+\exp{(\frac{E-E_{\rm
F}(T_{\rm e})}{k_{\rm B}T_{\rm e}}})}
\end{equation}
considering the total electron density $N_{\rm e}$ and using the
calculated density of states ${\rm DOS}(E)$ from
Ref.\,\onlinecite{LinPRB08}. To calculate the change on the Fermi
Level, we can limit the integration range from $E_{\rm F}^-=E_{\rm
F}(0)-10$\,eV to $E_{\rm F}^+=E_{\rm F}(0)+20$\,eV,

\begin{equation}\label{Fermi}
    \int_{E_{\rm
F}^-}^{E_{\rm F}(0)} {\rm DOS}(E) dE =
    \int_{E_{\rm
F}^-}^{E_{\rm F}^+}  \frac{{\rm
DOS}(E)dE}{1+\exp{\left(\frac{E-E_{\rm F}(T_{\rm e})}{k_{\rm
B}T_{\rm e}}\right)}}
\end{equation}

The limits of integrations in Eq.\,\ref{jointDOS} are calculated as in Ref.\,\onlinecite{RoseiPRB74,GuerresiPRB75} and
are $E^{\rm X}_{\rm min}=-20 k_{\rm B}T$, and
\begin{equation}\label{EmaxXa}
E^{\rm X}_{\rm
max}=\hbar\omega_6^X+(\hbar\omega-\hbar\omega_6^X-\hbar\omega_7^X)\frac{m^X_{d\parallel}}{m^X_{p\parallel}-m^X_{d\parallel}}
\end{equation}
if $\hbar\omega\leq\hbar\omega_6^X+\hbar\omega_6^7$, or
\begin{equation}\label{EmaxXb}
E^{\rm X}_{\rm
max}=\hbar\omega_6^X+(\hbar\omega-\hbar\omega_6^X-\hbar\omega_7^X)\frac{m^X_{d\perp}}{m^X_{p\perp}+m^X_{d\perp}}
\end{equation}
if $\hbar\omega>\hbar\omega_6^X+\hbar\omega_6^7$. Near the L point
they are
\begin{equation}\label{EmaxLa}
E^{\rm L}_{\rm
max}=-\hbar\omega_{4^-}^L+(\hbar\omega-\hbar\omega_0)\frac{m^L_{d\parallel}}{m^L_{d\parallel}-m^L_{p\parallel}}
\end{equation}
if $\hbar\omega\leq\hbar\omega_0$, or
\begin{equation}\label{EmaxLb}
E^{\rm L}_{\rm
max}=-\hbar\omega_{4^-}^L+(\hbar\omega-\hbar\omega_0)\frac{m^L_{d\perp}}{m^L_{d\perp}+m^L_{p\perp}}
\end{equation}
if $\hbar\omega>\hbar\omega_0$, and
\begin{eqnarray}\label{EminL}
E^{\rm L}_{\rm
min}&=&-\hbar\omega_{4^-}^L-\frac{\hbar^2}{2m^{L}_{p\parallel}}\left(\frac{\pi}{4a}\right)^2+\frac{m^{L}_{d\perp}}{m^{L}_{p\perp}+m^{L}_{d\perp}}\; \nonumber \\
&\times&\left[\frac{\hbar^2}{2}\left(\frac{1}{m^{L}_{p\parallel}}-\frac{1}{m^{L}_{d\parallel}}\right)\left(\frac{\pi}{4a}\right)^2+(\hbar\omega-\hbar\omega_0)\right]
\end{eqnarray}

The effective mass parameters and band gap parameters are taken from the Au band structure near the X and L
points\,\cite{ChristensenPRB71,GuerresiPRB75} and are $m_{p\parallel}^X=0.12\,$m$_0$, $m_{d\parallel}^X=0.91\,$m$_0$,
$m_{p\perp}^X=0.22\,$m$_0$, $m_{d\perp}^X=0.75\,$m$_0$, $m_{p\parallel}^L=0.25\,$m$_0$, $m_{p\perp}^L=0.23\,$m$_0$,
$m_{d\perp}^{L^{+}_{5+6}}=0.63\,$m$_0$, $m_{d\parallel}^{L^{+}_{5+6}}=0.7\,$m$_0$,
$m_{d\parallel}^{L^{+}_4}=0.57\,$m$_0$, $m_{d\perp}^{L^{+}_4}=0.49\,$m$_0$, $\hbar\omega_7^X=1.77$\,eV,
$\hbar\omega_6^X=1.47$\,eV, $\hbar\omega_{4^-}^L=0.72$\,eV, $\hbar\omega_{5+6}^L=1.45$\,eV,
$\hbar\omega_{4^+}^L=2.6$\,eV, and the lattice parameter is $a=4.08\times10^{-10}$\,m.

To fit the measured $\epsilon^{\rm b}_{2}$ at 300\,K by Johnson and Christy\,\cite{JohnsonPRB72} we use the constants
$A_{\rm X}$, $A_{\rm L^{+}_{5+6}}$, $A_{\rm L^{+}_{4}}$ as fitting parameters similar to
Ref.\,\onlinecite{GuerresiPRB75}, and obtain $|P_{\rm X}/P_{\rm L^{+}_{5+6}}|^2=0.26$, $|P_{\rm X}/P_{\rm
L^{+}_{4}}|^2=0.67$. One also needs to account for a broadening of $\epsilon^{\rm b}_{2}$ in order to fit the
low-energy tail experimentally observed. Instead of introducing an effective temperature as fit
parameter\,\cite{GuerresiPRB75}, we consider the Drude broadening by convoluting the expression calculated according to
Eq.\,\ref{epsilon2} with the function
\begin{equation} B(\hbar\omega)=\frac{1.76}{2b\Gamma}\cosh^{-2}\left(\frac{1.76\hbar\omega}{b\Gamma}\right) \end{equation}
which has an integral normalized to 1 and a full-width at half-maximum $b\Gamma$ proportional to the Drude broadening.
We use
\begin{equation}\label{broadening}
    b\Gamma=5\hbar\left[\gamma_0 + \left(b_{\rm ee}(T_{\rm
e}^2-T_0^2)+ b_{\rm ep}(T_{\rm l}-T_0)\right)(\hbar\omega_{\rm exc})^2\right]
\end{equation}
which fits well the data of Johnson and Christy\,\cite{JohnsonPRB72}
at 300\,K and accounts for additional broadenings due to
electron-electron and electron-phonon scattering processes in the
presence of a non-equilibrium electron and lattice temperature on
the same ground as for the Drude part.

\subsection{Coherent phonon oscillations}

The visibility of coherent phonon oscillations in $\Delta\widetilde{\epsilon_{\rm R}}$ is described by the change of
the plasma frequency due to the changing volume of the NP during the breathing oscillation, driven by the non-adiabatic
thermal expansion of the NP during the thermalization with the lattice in the first picoseconds:
\begin{equation}\label{coherentphonons}
    \omega_{\rm p}=\frac{\omega_{\rm p0}}{\sqrt{\overline{\alpha}(T_{\rm l}-T_0)\left(1+C\cos{\left(\frac{2\pi\tau}{T_{\rm o}}+\phi\right)}e^{-\frac{\tau}{T_{\rm d}}}\right)+1}}
\end{equation}
with the bulk volumetric expansion coefficient $\overline{\alpha}=4.5\times10^{-5}$\,K$^{-1}$. To fit the oscillations
measured on the 38\,nm NP shown in the paper (inset of Fig.\,2) we used $C=0.75$, $T_{\rm o}=12.8$\,ps, $T_{\rm
d}=80$\,ps, and $\phi=-3.33$. The oscillation period is the period of the fundamental breathing mode in a spherically
symmetric NP\,\cite{HartlandChemRev11}, from which we deduce a particle diameter of 38\,nm following the values in
Ref.\,\onlinecite{VanDijkPRL05}. The quantitative agreement of the oscillation amplitude using a factor of $C=0.75$ is
supporting this interpretation, since for an instantaneous lattice heating we would expect $C=1$, and for a realistic
heating time of a few picoseconds this value is expected to be somewhat reduced.

\section{FWM dynamics for large nanoparticles}

\renewcommand{\thefigure}{\Roman{figure}}
\begin{figure}[t!]
\includegraphics*[width=7cm]{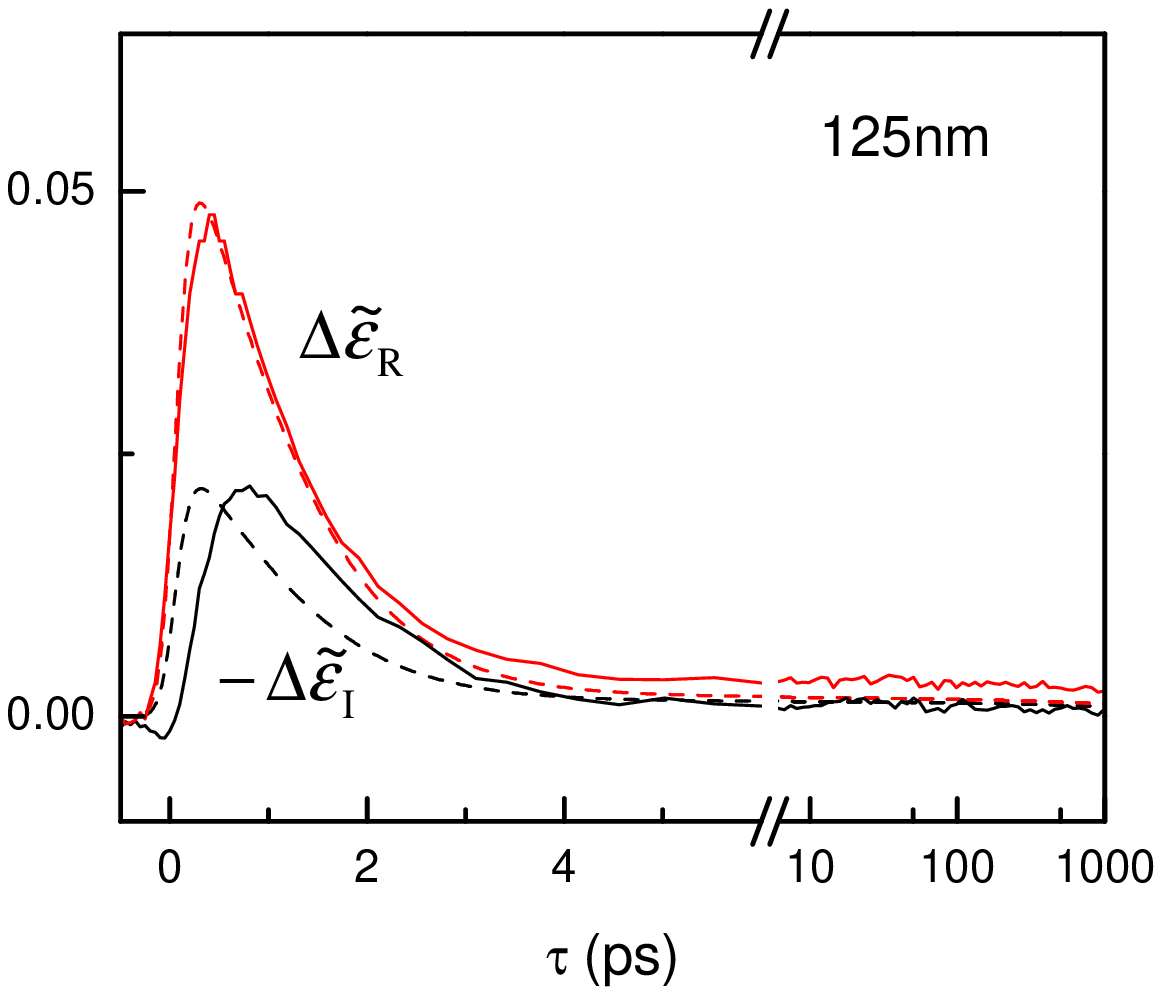}
\caption{Transient changes of the real
($\Delta\widetilde{\epsilon_{\rm R}}$) (red) and imaginary part
($\Delta\widetilde{\epsilon_{\rm I}}$) (black) of the dielectric
function of a single 125\,nm gold nanoparticle. Pump (probe) fluence
is 0.65\,J/m$^{2}$ (0.05\,J/m$^{2}$). Dashed lines are calculations
as described in the text.\label{Fig1SM}}
\end{figure}

The SPR of a GNP in a dielectric surrounding can be described by the dipole approximation for a GNP diameters of less
than about 40\,nm. For larger sizes retardation effects become significant, resulting in a redshift of the extinction
peak with increasing size. To verify the suitability of our model to describe large NP sizes, we have investigated the
dynamics of $\Delta\tilde{\epsilon}$ for a 125\,nm GNP excited and probed at 550\,nm (solid lines in
Fig.\,\ref{Fig1SM}). The size of the NP has been estimated by the oscillation period measured in the phase part of the
signal $P$ (see previous section). A major difference with respect to the smaller NP sizes investigated is that $P$ has
a positive sign. Within our model in the dipole approximation, we can reproduce the position of the absorption cross
section (584\,nm) obtained including retardation using Mie's theory\,\cite{vanDijkPhD} by reducing the plasma frequency
to 7.965\,eV. However when modeling the FWM dynamics (see dashed lines in Fig.\,\ref{Fig1SM}, the particle-dependent
parameters used in these simulations are $G_{\rm ep}=2.5\times10^{-5}$\,W/K and $g=1.8$) we can reproduce the overall
dynamics only qualitatively. We do correctly simulate the signs of $\Delta\tilde{\epsilon}_{\rm R,I}$ and their
relative amplitude, with the positive sign of $P$ being a consequence of probing on the blue side of the SPR. However,
the delayed rise of $\Delta\tilde{\epsilon_{\rm I}}$ observed in the experiment is not reproduced by the model.
Moreover the proportionality factor $\xi$ between $A$ ($P$) and $\Delta\tilde{\epsilon_{\rm I}}$
($\Delta\tilde{\epsilon_{\rm R}}$) does not scale with the NP volume. We attribute these discrepancies to the failure
of the dipole approximation.

\section{Validity of the Energy Transfer in the two-temperature model}

\renewcommand{\thefigure}{\Roman{figure}}
\begin{figure}
\includegraphics*[width=8cm]{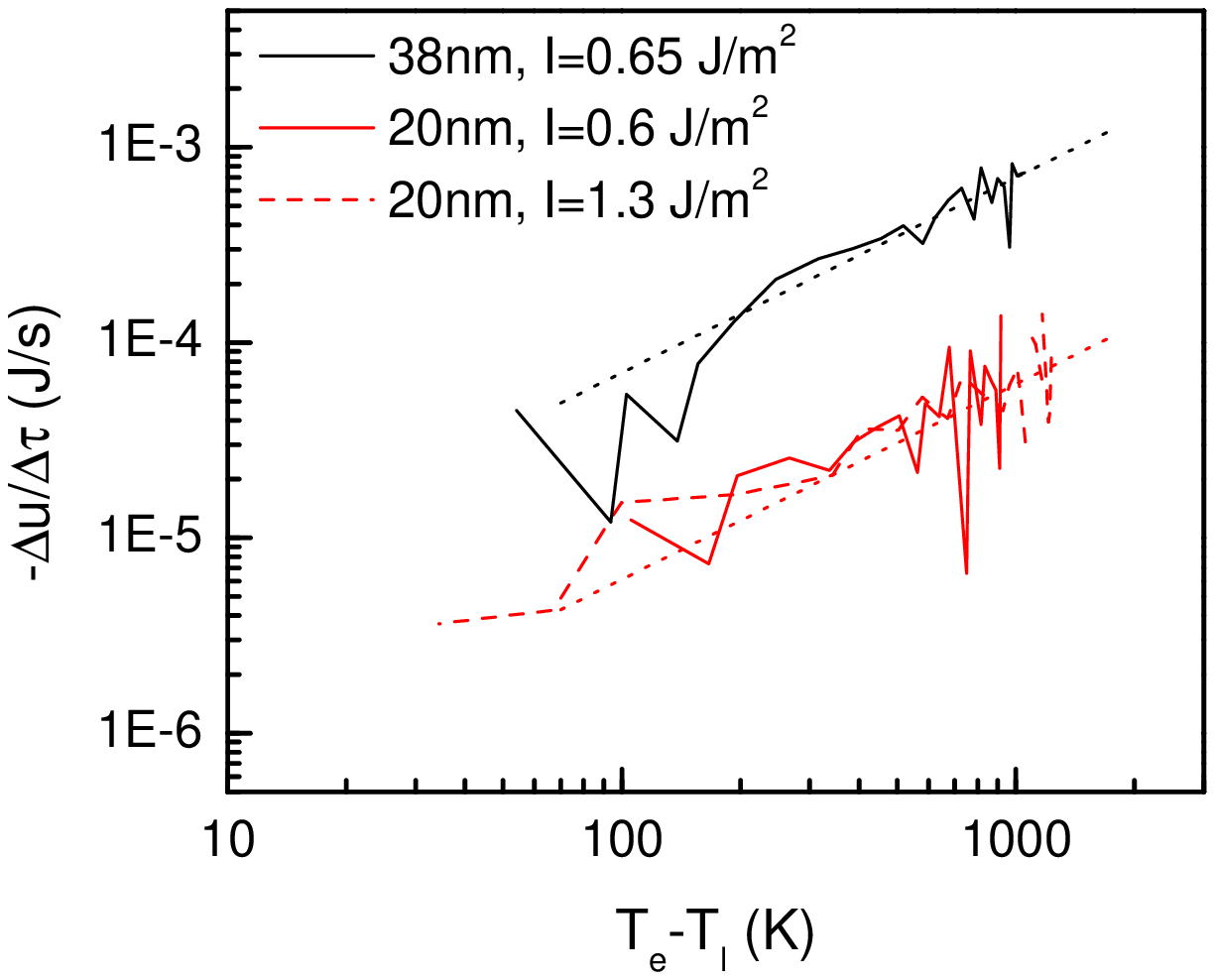}
\caption{Electron energy transfer rate as a function of $T_{\rm
e}-T_{\rm l}$ for 38\,nm NP with excitation intensity of
0.65\,J/m$^2$ (black line) at 550\,nm and for a 20\,nm NP with
excitation intensity of 0.60 (red solid line) and 1.3\,J/m$^2$ (red
dashed line) at 550\,nm. The dotted lines represent a linear
dependence with slopes which are proportional to the electron-phonon
coupling constant $G_{\rm ep}$.\label{Fig2SM}}
\end{figure}

To demonstrate the validity of the two-temperature model we have
investigated the dependence of the energy transfer rate from the
electrons to the lattice $\Delta u_{\rm el}/\Delta\tau$ on the
difference between the electron and lattice temperature $T_{\rm
e}-T_{\rm l}$. All these quantities can be retrieved directly from
the FWM signal when we assume that the total amplitude $E_{\rm
FWM}=\sqrt{A^2+P^2}$ of the field can be approximated as

\begin{equation}\label{totalFWM}
E_{\rm FWM}=a\left(T_{\rm e}^2-T_0^2\right)+b\left(T_{\rm
l}-T_0\right).
\end{equation}
The electron and lattice temperature can then be determined from the FWM amplitude considering the energy conservation
within the NP, which holds for delay times much shorter than the typical thermalization time with the environment of
some 100\,ps,
\begin{equation}\label{totalen}
u=\frac{\gamma T_{\rm e}}{2}\left(T_{\rm e}^2-T_0^2\right)+C_{\rm
l}\left(T_{\rm l}-T_0\right)
\end{equation}
leading to
\begin{equation}\label{TevsFWM}
T_{\rm e}=\left(\frac{E_{\rm FWM}- b u/C_{\rm l}}{a-b\gamma/(2C_{\rm
l})}+T_0^2\right)^{1/2}
\end{equation}
and
\begin{equation}\label{TlvsFWM}
T_{\rm l}=\frac{1}{C_{\rm l}}\left(u-\frac{\gamma}{2}\left(T_{\rm
e}^2-T_0^2\right)\right)+T_0.
\end{equation}
The values of $a$ and $b$ can be found by comparing the dynamics of $T_{\rm e}$ calculated with the two-temperature
model and the value of the electron temperature obtained by Eq.\,\ref{TevsFWM}. The value of the energy $u$ has been
estimated from the incident energy flux and the nanoparticle absorption cross section. Once the values of $a$ and $b$
are determined, the electron energy transfer rate can be calculated in a finite difference scheme between the $i$-th
and $i+1$-th point of the measured FWM amplitude dynamics:
\begin{equation}\label{elentransfer}
\frac{\Delta u}{\Delta\tau}=\frac{\gamma}{2}\frac{T_{{\rm
e}_{i+1}}^2-T_{{\rm e}_{i}}^2}{\tau_{i+1}-\tau_i}.
\end{equation}
The values of $-\Delta u/\Delta\tau$ as a function of the
electron-lattice temperature difference $T_{\rm e}-T_{\rm l}$ were
plotted in Fig.\,\ref{Fig2SM} for a 38\,nm NP with excitation
intensity of 0.65\,J/m$^2$ at 550\,nm and for a 20\,nm NP with
excitation intensity of 0.60 and 1.3\,J/m$^2$ at 550\,nm. The data
refer to a delay range of 0.7-5\,ps, i.e. in the cooling regime from
the electron gas to the lattice. Within the experimental errors, the
energy transfer rate depends linearly on $T_{\rm e}-T_{\rm l}$,
confirming the validity of the two temperature model. As expected,
the amplitude of the slope does not depend on the excitation
intensity and it is proportional to the value of the electron-phonon
coupling constant $G_{\rm ep}$.

\section{Ratiometric detection of the signal phase}

\renewcommand{\thefigure}{\Roman{figure}}
\begin{figure}[t!]
\includegraphics*[width=8cm]{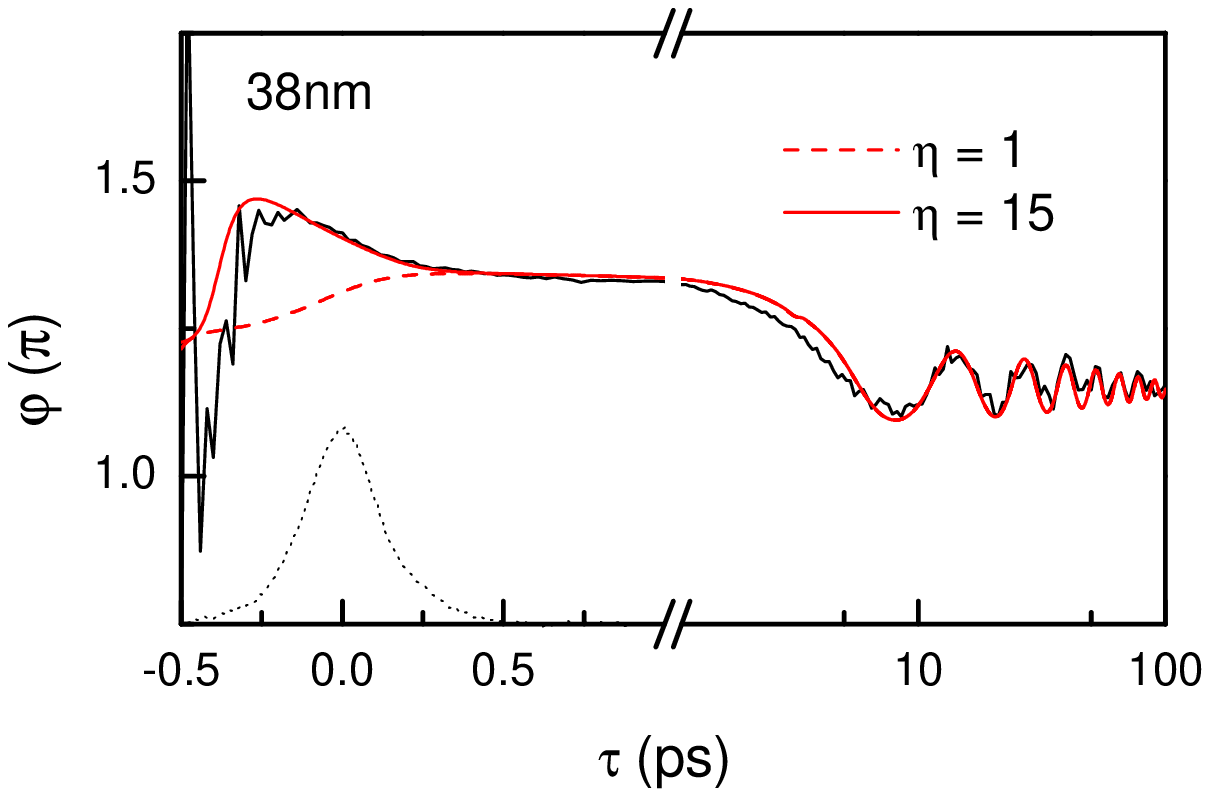}
\caption{Transient change of the phase of $\Delta\tilde{\epsilon}$
from the ratio between its imaginary and real part, measured on the
38\,nm NP for an excitation and detection wavelength of 550\,nm. Red
dashed line is a simulation assuming the parameter $\eta=1$ in the
shift of the plasma frequency (Eq.\,\ref{plasmafreq}). The solid
line assumes $\eta=15$ as indicated. The black dotted line is the
pulse intensity autocorrelation measured in the experiment.
\label{Fig3SM}}
\end{figure}

An intrinsic ratiometric way to represent the measured FWM signal is to plot the phase $\varphi$ of
$\Delta\widetilde{\epsilon}=|\Delta\widetilde{\epsilon}|\exp(i\varphi)$ from the ratio between its imaginary and real
part (hence $A$ and $P$). This is shown in Fig.\,\ref{Fig3SM} for the 38\,nm NP at 550\,nm. Noticeably, the visibility
of the coherent phonon oscillations is improved in $\varphi$ since it is independent of decay dynamics common to real
and imaginary part of $\Delta\widetilde{\epsilon}$, while a frequency shift in the SPR changes $\varphi$.

We point out that our simulations include the effect of the change
in the plasma frequency from the excess electron density calculated
as:

\begin{equation}\label{plasmafreq}
\omega_{\rm p}=\omega_{\rm p}^0 \sqrt{\left(1+\eta\frac{n_{\rm X}^{\rm nth}+n_{\rm L}^{\rm nth}}{n_{\rm e}}\right)}.
\end{equation}

The presence of an excess electron density equal to the density of
holes generated via the interband transitions is corresponding to
$\eta=1$. Indeed all simulations shown in the paper for the dynamics
of $\Delta\tilde{\epsilon_{\rm I}}$ and $\Delta\tilde{\epsilon_{\rm
R}}$ are performed using $\eta=1$. However, in the dynamics of the
signal phase we notice that the initial dynamics on the time scale
of the excess density lifetime (i.e. for $\tau\leq0.5$\,ps) is
better reproduced with $\eta=15$ (red solid line in
Fig.\,\ref{Fig3SM}). A possible physical explanation for $\eta=15$
is the influence of the hole density and the creation of carriers in
specific points of the Brillouin zone with effective masses which
are different from the average effective mass $m_0$ used in the
definition of the plasma frequency.

\renewcommand{\thefigure}{\Roman{figure}}
\begin{figure}[t!]
\includegraphics*[width=8cm]{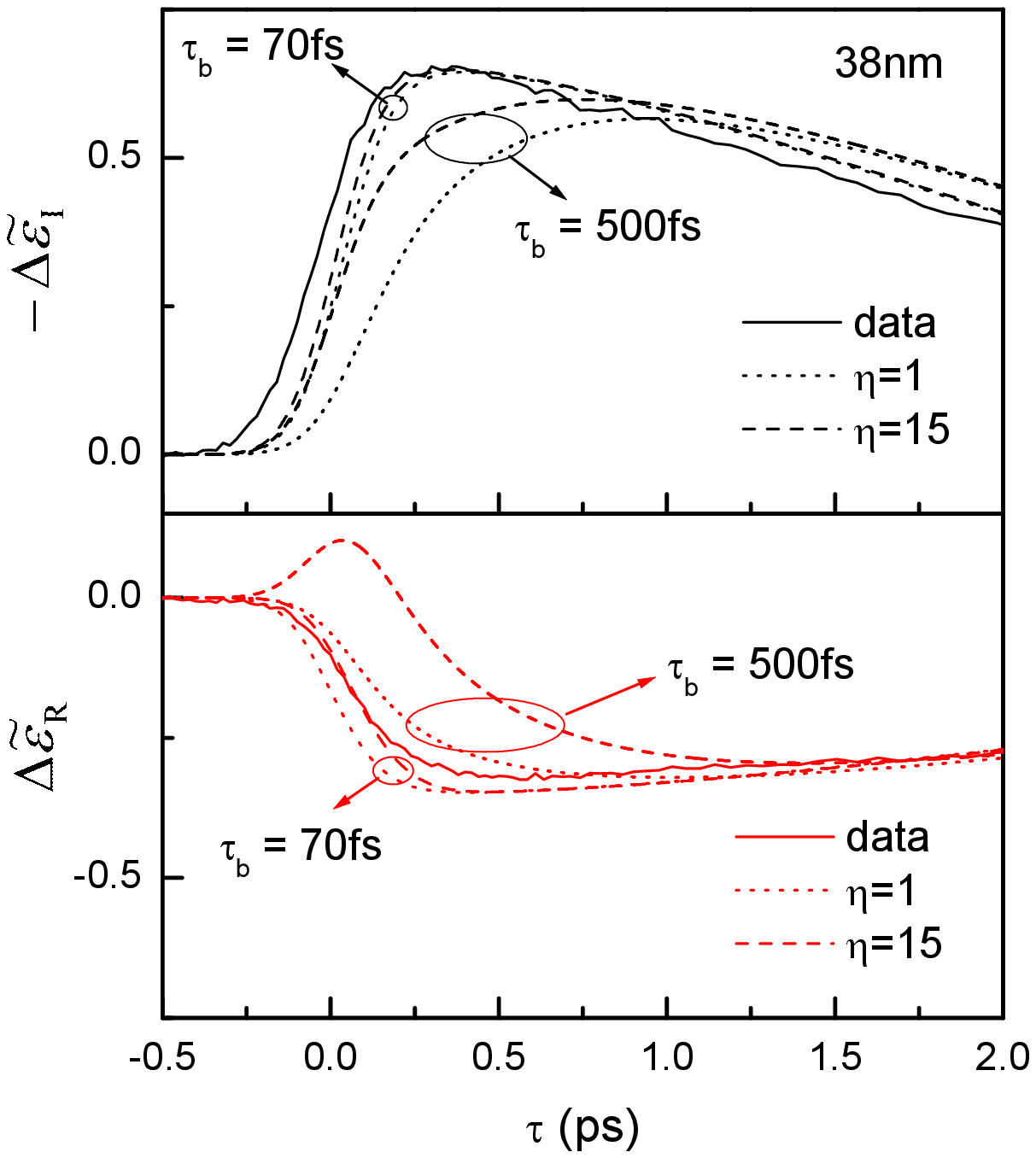}
\caption{Transient change of the real and imaginary part of
$\Delta\tilde{\epsilon}$ measured on the 38\,nm NP for an excitation
and detection wavelength of 550\,nm, with simulations assuming
$\eta=1$ (dotted lines) and $\eta=15$ (dashed lines) and for
$\tau_{\rm b}=70$\,fs or $\tau_{\rm b}=500$\,fs as
indicated.\label{Fig4SM}}
\end{figure}

When looking at the simulations of $\Delta\tilde{\epsilon_{\rm I}}$ and $\Delta\tilde{\epsilon_{\rm R}}$  at 550\,nm
(see Fig.\,\ref{Fig4SM}) the effect of $\eta$ manifests as a small shift in the initial rise which goes in opposite
directions for the real and imaginary part of $\Delta\tilde{\epsilon}$, namely increasing $\eta$ shifts
$\Delta\tilde{\epsilon_{\rm I}}$ toward earlier times and $\Delta\tilde{\epsilon_{\rm R}}$ toward later times. This
effect however does not modify significantly the initial rise time, which instead is determined by the d-state Auger
thermalization time $\tau_{\rm b}=70$\,fs. To elucidate this point we show in Fig.\,\ref{Fig4SM} the effect of changing
$\tau_{\rm b}$ to 500\,fs (equal to the free electron thermalization time constant $\tau_{\rm f}$) for both $\eta=1$
and $\eta=15$. A longer $\tau_{\rm b}$ clearly shows as a slower rise time dynamics which cannot be compensated by
changing $\eta$ and does not agree with the measurements.



%